\documentclass[11pt]{article}
\usepackage{blois,epsfig,graphicx}

\def\Journal#1#2#3#4{{#1} {\bf #2}, #3 (#4)}

\def\be{\begin{equation}}
\def\ee{\end{equation}}
\def\bea{\begin{eqnarray}}
\def\eea{\end{eqnarray}}

\begin{document}
\vspace*{4cm}
\title{MIGALE : A MULTIPARAMETRIC VIRTUAL INSTRUMENT TO STUDY GALAXY EVOLUTION}

\author{ G. THEUREAU }

\address{GEPI, Paris Observatory, and LPCE, CNRS-ORLEANS, \\
3A avenue de la Recherche Scientifique, 45071 Orl\'eans, France}

\maketitle
\abstracts{
Galaxy evolution is a complex process where both the inner evolution of stellar population, gas and dust,
and the external effects, like interactions and exchanges with the environment, have to be taken into account.
It has been fundamental in the last years to be able to build and use homogeneous
catalogues both in the local and the far universe. The observation of galaxy morphology and kinematics
as a function of the redshift is indeed necessary to disentangle the various galaxy formation
and evolution scenarios.
Some years ago, the hyperleda extragalactic database was designed to study the local
universe from the point of view of both stellar populations and galaxies kinematics and dynamics. 
Today it contains homogeneous data for about 3 millions of galaxies,
with for each up to 80 astrophysical parameters available.
We will describe here the MIGALE project which emcompasses the HyperLeda databases plus a
series of tools developed to study the dynamical, chemical and morphological
evolution of galaxies. It will include, in particular, methods to analyse the
GIRAFFE cosmological fields (IFU spectroscopy) and compare them with the Local
Universe. }

\section{Scientific context}

The formation of the galaxies and the stars that they are made of is one of the
great puzzles of modern astrophysics. The first surveys (CFRS, HDF and Keck)
indicate that the peak of stellar formation should be between $z=$1 and $z=$1.5
(Madau et al, 1996). HST observations show that at large distances ($z>$ 0.7)
the galaxies with strong stellar formation ($>$ 10$M_{\odot}/year$) are either
interacting systems (detected by ISO \& VLA; Flores et al, 1999) or compact galaxies
(Guzman et al, 1997), with the comobile density of disks decreasing very progressively
with $z$ (Lilly et al, 1998). However, numerous fundamental questions remain
unsolved, for example:
What is the quantity of mass accreted by coalescence when the galaxy forms ?
What is the importance of galaxy interactions ?
How do the numerous population of compact galaxies evolves from large $z$ ?
How do we explain the stability and evolution of galaxy disks ? 
What is the role of spiral arms and bars ?
How is redistributed the angular momentum that was acquired by tidal interactions between halos
and how it is related to the disks sizes ?

Therefore, the observation of galaxy morphology and kinematics during
their evolution and as a function of the redshift is necessary
to disentangle the various formation and evolution scenarios.
It is also fundamental to understand how stellar populations evolve and what are the associated 
feedback processes which could delay the gas from cooling and its collapse into disks.

\section{The database and the virtual instrumentation}

The statistical observation of galaxy kinematics appears to be the only means
to shed some light on or even perhaps solve the puzzle of the angular momentum.
The building of template samples in the local universe is also fundamental in
understanding the properties of distant galaxies. For this reason,
we have decided to group together the two multiparametric and multiwavelength databases
HYPERLEDA (local universe) and GIRAFFE (cosmological fields at intermediate redshifts)
in the frame of the same structure of virtual instrumentation.
This statistical approach will allow us to understand, in particular, how the Hubble sequence
was formed from distant galaxies, which have a less structured morphology compared to present
day galaxies, and to identify the mechanism that triggers the intense stellar formation.

\subsection{What is MIGALE ?}

MIGALE (Multi-parametric virtual Instrument for the study of GALaxy Evolution) consists
of a central node, hosting in particular the LEDA catalogue of homogenized parameters and
a set of catalogues containing raw data tables (updated from litterature and 
some large surveys), 
some links towards associated observational projects and archives, some data mining tools 
to point towards external archives and on-the-fly treatment softwares developped by various 
associated scientific teams and which allows us to manipulate the extracted images or spectra.

MIGALE is then based on a multiparametric approach cross-identifying, homogenizing and 
combining data from many different sources, in particular wide field (or whole sky) samples
such as DSS, DENIS, 2MASS, SDSS or 2dF. It allows us to manage and acces the data downstream 
of the telescope's archives, develope new analysis tools and distributes them either as packages 
or through web access.

From the point of view of an external user, one can for example:
\begin{itemize}
\item select samples, specifying simultaneously various constraints on numerous physical parameters: e.g.
select all edge-on Sb galaxies brighter than 14. in B and having a 21-cm line measurement
\item do statistical studies on large samples: e.g. study the Tully-Fisher or Fundamental plane scaling laws
as a function of morphology or redshift; study the stellar population characteristics as a function of environment
\item access and process template data: e.g. extract, rescale, convolve and combine some available high resolution 
stellar spectra to mimic your own low or medium resolution observed spectrum
\item access and process data from several large surveys: e.g. extract, filter and combine subimages from the 
DENIS and the 2MASS surveys 
\item use virtual instruments:  e.g. create a synthetic galaxy spectrum according to some stellar population characteristics
\end{itemize}

\subsection{The historical LEDA catalogue}

The LEDA catalogue was built from an exhaustive compilation of litterature and thanks to
its association with some large observational projects (DENIS, KLUN+, DSS...). It has grown
from 73,000 galaxies in 1989 to about 3 millions today (see e.g. Paturel et al 2003a). 
It was used in 1993 to create the RC3 catalogue. 
LEDA maintains, in collaboration with NED, a general index of galaxies, 
characterised by the pgc number. It also: 1) provides on  the fly data homogenization 
and correction, 2) displays on line charts and DSS images, 3) produces an homogenized catalogue 
of 80 astrophysical mean parameters per galaxy and 4) permits SQL access
to all the stored data. The statistics for a magnitude complete sample up to $B$=18 is given in
table 1.

\begin{table}[t]
\caption{LEDA statistics for a magnitude complete sample up to $B$=18 ($\sim$1.2 millions of galaxies)}
\vspace{0.4cm}
\begin{center}
\begin{tabular}{ll}
parameters & percentage \\ \hline
magnitudes (BIJHK) & 74\%  \\
dimensions, axis ratios and pos. angle &  74\% \\
morphology &  11\% \\ 
rotational velocity &  2\% \\ 
central velocity disp. &  $<$ 1\%  \\ \hline 
\end{tabular}
\end{center}
\end{table}


\subsection{The Hypercat contribution}

The Hypercat concepts are at the heart of the LEDA's transformation 
towards HYPERLEDA's datamining framework and now MIGALE.
In addition to the principle of a network database involving 
several nodes driven by different scientific teams, the Hypercat team
has built up the cornerstone of the present project, the PLEINPOT 
package, which contains the infrastructure and all the facilities
from which the database and the analysis tools are constructed.
The package contains many modules related to interface and data acces,
statistics, astronomy and image analysis. Among these on finds the
following possibilities: flat-field correction, wavelength calibration,
flux calibration, flux normalization, extraction of a subimage either from
its coordinates or its name, projection (2D $\rightarrow$ 1D), substraction
of sky background, removal of cosmics, mask or filter application...
In addition to the LEDA catalogues, the HYPERLEDA site already provides the following 
facilities. The pixel server distributes or give access to the images from DSS1, DSS2, 
DENIS (I and J bands) and 2-MASS (J, H and K bands). The ELODIE library of stellar 
high resolution spectra is available on-line (1970 spectra for 1200 stars with a 
resolution of 10,000 and a wavelength range of 410--480 nm) and the PEGASE simulator 
(evolutive spectral synthesis, Fioc \& Rocca 1997) is also interfaced.

\subsection{The HI catalogue and the KLUN+ survey}

The aim of the KLUN+ project is to collect an homogeneous Tully-Fisher sample of 
20,000 field spiral galaxies distributed over the whole sky. This sample is gathered
both from an exhaustive compilation of the litterature and from a large complementary 
programme of 21-cm line observations.  From June 2001, our "KLUN+" programme has 
been accepted as the Cosmological Key-project of the refurbished Nan\c cay 
radiotelescope and is allocated about 25\% of the telescope time. 
The on-line HI archive give acces to the 21-cm line profile for $\sim$ 4500 galaxies.
Our last HI data compilation contains today 16,666 galaxies, from 611 references, among them 5263 
galaxies observed with the Nan\c cay antenna (Paturel et al 2003b, Theureau et al 2003).
This calalogue has permitted in particular a detailled investigation into local universe peculiar 
velocity field and large scale structures (Hanski, Theureau \& Paturel 2003 and this conference) 

\subsection{The CAI component}

The Center of Image Analysis (CAI) manages the MAMA multibank microdensitometer and produces
high resolution scan images from photographic plates.  Its resolution is 0.6 arcsec, thus much 
better than e.g. DSS1 and DSS2. The whole atlas SRC-J and ESO-R covering the southern sky as 
well as their corresponding catalogues will be made available in 2004. 

\subsection{GIRAFFE and internal dynamics}

The Giraffe database is devoted to the exploitation of the stellar and extragalactic 
spectroscopic data provided  by the GIRAFFE spectrograph. This database will supply the 
users with : the spectra and the deduced parameters (e.g. in total up to 30,000 spectra
in the 35 nights of guaranted time), the wide field images resulting from the preparation 
of GIRAFFE observations with their astrometry, the associated images at high resolution
and the photometric catalogues.

GIRAFFE cosmological fields observations will produce in particular high resolution
galaxy spectra up to $z$=1 providing detailed emission line profiles ($H\alpha$, $H\beta$,
$[OIII]$ ...) and, in the IFU or ARGUS modes, allowing very accurate galaxy internal 
velocity field studies (Fig. 1). 
           

\begin{figure}
\includegraphics[width=14cm]{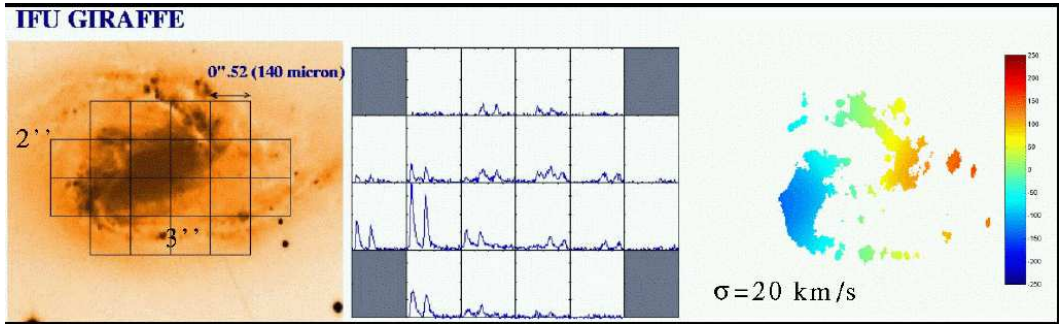}
\caption{From left to right: IFU spectrograph mode superimposed on the galaxy image, measured
set of spectra from each cell/fiber, reconstructed 3D velocity field (Cayatte, Chemin \& Flores 2003)
}
\end{figure}

\section*{Acknowledgments}
This project is the work of the whole group and I thank all the collaborators of the MIGALE project, 
in particular the members of the project group: P.Prugniel (P.I), F.Tajahmady (P.M), H.Flores, J.Guibert,
I.Jegouzo, F.Royer and J.V\'etois. Also acknowledged are the institutes supporting the project: the GEPI
(Paris Observatory), the CRAL (Lyon Observatory) and the PNG (Programe National Galaxies, CNRS, 
France).

\end{document}